**Graphical Abstract (GA)**

**GA Figure:**

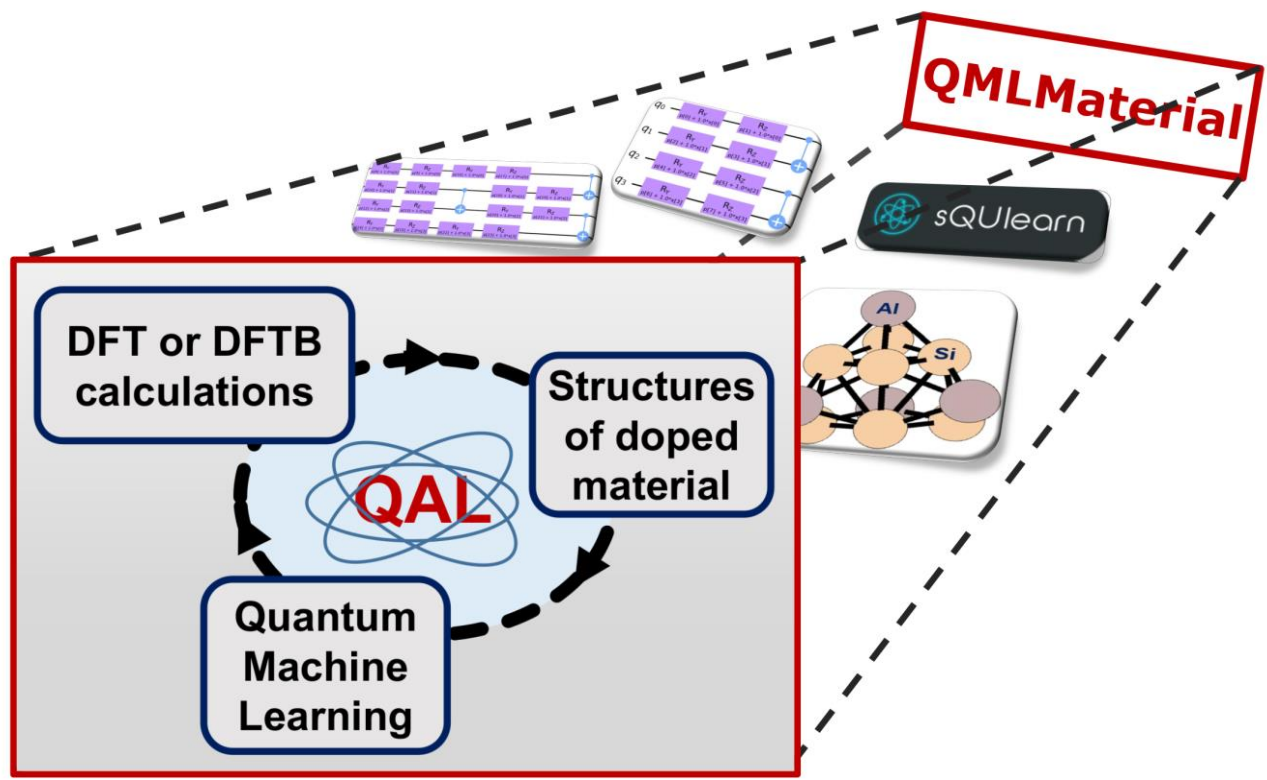

**GA Text:** A quantum active learning method (QAL) for automatic structural determination of doped materials has been developed and implemented in the QMLMaterial software. QAL uses quantum circuits for data encoding to create quantum machine learning models on-the-fly.

# Quantum Active Learning for Structural Determination of Doped Nanoparticles – a Case Study of 4Al@Si$_{11}$

*Maicon Pierre Lourenço[a*], Mosayeb Naseri[b], Lizandra Barrios Herrera[b], Hadi Zadeh-Haghighi[c], Daya Gaur[d], Christoph Simon[c] and Dennis R. Salahub[b]*

[a]*Departamento de Química e Física – Centro de Ciências Exatas, Naturais e da Saúde – CCENS – Universidade Federal do Espírito Santo, 29500-000, Alegre, Espírito Santo, Brasil.*

[b]*Department of Chemistry, Department of Physics and Astronomy, CMS Centre for Molecular Simulation, IQST Institute for Quantum Science and Technology, Quantum Alberta, QHA Quantum Horizons Alberta, University of Calgary, 2500 University Drive NW, Calgary, AB, T2N 1N4, Canada.*

[c]*Department of Physics and Astronomy, Institute for Quantum Science and Technology, Quantum Alberta, and Hotchkiss Brain Institute, University of Calgary, Calgary, AB T2N 1N4, Canada.*

[d]*Department of Computer Science, University of Lethbridge, 4401 University Dr. West Lethbridge, AB T1K 3M4, Canada.*

*maiconpl01@gmail.com

ORCID ID: Maicon Pierre Lourenço (https://orcid.org/0000-0002-0110-8318); Hadi Zadeh-Haghighi (https://orcid.org/0000-0003-3380-9925); Dennis R. Salahub (https://orcid.org/0000-0002-9848-3762).

**Abstract**

Active learning (AL) has been widely applied in chemistry and materials science. In this work we propose a quantum active learning (QAL) method for automatic structural determination of doped nanoparticles, where quantum machine learning (QML) models for regression are used iteratively to indicate new structures to be calculated by DFT or DFTB and this new data acquisition is used to retrain the QML models. The QAL method is implemented in the Quantum Machine Learning Software/Agent for Material Design and Discovery (QMLMaterial), whose aim is using an artificial agent (defined by QML regression algorithms) that chooses the next doped configuration to be calculated that has a higher probability of finding the optimum structure. The QAL uses a quantum Gaussian process with a fidelity quantum kernel as well as the projected quantum kernel and different quantum circuits. For comparison, classical AL was used with a classical Gaussian process with different classical kernels. The presented QAL method was applied in the structural determination of doped $Si_{11}$ with 4 Al ($4Al@Si_{11}$) and the results indicate the QAL method is able to find the optimum $4Al@Si_{11}$ structure. The aim of this work is to present the QAL method – formulated in a noise-free quantum computing framework – for automatic structural determination of doped nanoparticles and materials defects.

1. Introduction

Machine learning is a branch of artificial intelligence that enables computers to learn from data and improve their performance on tasks without explicit programming. It uses algorithms to identify patterns and make predictions based on input data. While machine learning is a powerful approach in many scenarios, it also faces two main challenges: insufficient training data and the complexities associated with high-dimensional data.

To address the first mentioned challenge in machine learning, the active learning method can be used. It allows regression models to selectively query and obtain labels for the most informative data points, enhancing learning efficiency with fewer labeled examples. Therefore, practitioners can effectively mitigate the limitations posed by limited training data, improving model performance and adaptability across various applications. For instance, classical active learning (AL) has been widely applied in chemistry and materials science for structural determination of nanoparticles[1,2], solids and nanoparticles with doped sites and point defects[3], adsorption on surfaces of solids and nanoparticles[4,5], materials design such as perovskites[6].

To tackle the challenges of high dimensionality in machine learning, several methods can be used. Dimensionality reduction techniques like Principal Component Analysis[7] (PCA) and feature selection methods simplify datasets while preserving their structure. Although these methods are effective in many cases, they still face challenges in specific applications, such as drug discovery and other complex domains, where high dimensionality can hinder performance.

Quantum machine learning[8,9] (QML) can be considered as a promising alternative for addressing the challenges of high dimensionality in data. By leveraging the principles of quantum computing, such as superposition and entanglement, QML can process information in ways that classical algorithms cannot. This enables quantum algorithms to explore multiple possibilities simultaneously, making them particularly well-suited for handling complex, high-dimensional datasets.

QML is a rapidly progressing field with potential to bring quantum advantages to real-world problems[10]. A practical approach for near-term applications involves creating learning models based on parametrized quantum circuits[11]. These quantum models have demonstrated notable performance in benchmark tasks, such as supervised learning with quantum-enhanced feature spaces[12], differentiable learning of quantum circuit Born machines[13], and reinforcement learning[14]. These tasks have been tested in both computer simulations, where circuits were optimized using classical resources[15], and on actual quantum hardware. Real quantum implementations include supervised learning on quantum-enhanced feature spaces[12], training on hybrid quantum computers[16], handling high-dimensional data on noisy quantum processors[17], and processing large datasets with randomized measurements[18]. Their success in these areas suggests they could tackle complex problems beyond the reach of classical algorithms, such as predicting ground state properties of highly-interacting quantum systems[19].

Given the challenges of insufficient data and high dimensionality in classical machine learning and the mentioned schemes to tackle these issues, it seems reasonable to combine the concepts of active learning with quantum machine learning to create a novel approach known as Quantum Active Learning[20, 21] (QAL). By integrating the selective data querying capabilities of classical active learning with the advanced processing power of quantum algorithms, QAL has the potential to address both challenges simultaneously. This hybrid approach can efficiently identify the most informative data points for labeling while leveraging quantum computing's ability to handle high-dimensional spaces.

Recently, we proposed Quantum Active Learning[20] (QAL) utilizing either a quantum support vector regressor[22] (QSVR) or a quantum Gaussian process regressor[23] (QGPR), employing various quantum kernels and feature maps to guide new experiments or computations for optimization of perovskite properties such as piezoelectric coefficient, band gap, energy storage and the structure optimization of a doped nanoparticle ($3Al@Si_{11}$) together with the optimum spin multiplicity with as little data as possible. Our results revealed that the QAL method improved the searches in most cases.

In this work, we apply our new artificial intelligence approach based on quantum machine learning and active learning – a QAL approach – to guide new computations for automatic structural determination of doped nanoparticles[2], aimed at finding the global minimum structure with as few configurations calculated or visited as possible. QAL is based on quantum supervised machine learning (ML), in particular quantum Gaussian process[23] (GPR). More data is acquired during the QAL cycle and the QML regression is improved, increasing the probability of finding the optimum structure with the fewest possible new DFT[24] or DFTB[25] computations. The purpose of this work is to present the QAL method, formulated in a noise-free quantum computing framework, for automatic structural determination of doped nanoparticles and materials defects.

The QAL method is implemented in the Quantum Machine Learning Software/Agent for Material Design and Discovery[26] (QMLMaterial). The QAL method was applied in the structural determination of $Si_{11}$ nanoparticle doped with Al ($4Al@Si_{11}$)[2, 27]. The QAL results were compared to classical AL using classical GPR[28]. The QAL method as implemented and formulated can be applied to other doped nanoparticles[20] (e.g: different sizes) as well solids[20]. Note that the QAL method implemented in QMLMaterial uses quantum information for building QML models as well as quantum chemistry methods (e.g. DFTB) to compute the $4Al@Si_{11}$ homotops energies. This can be seen as a "QQ" method, i.e.: quantum information (Q) for inferences and quantum calculations (Q) for molecules.

Fig. 1A shows the putative global minimum (GM) found for the $4Al@Si_{11}$ doped nanoparticle and the energy distribution for the 330 $4Al@Si_{11}$ homotops obtained by DFTB calculations[2] (Fig. 1B). This makes a good database to benchmark the proposed QAL for structural search of doped nanoparticles. Note that the vertical bar in Fig. 1B marks the GM energy (Fig. 1A). The QAL study was designed in order to start from an observed homotop population (i.e.: calculated by DFT or DFTB) whose energies are far away from the GM (e.g.: -12.2400 Hartree). Subsequently, a "quantum agent" (built from a QML model using the observed data) for decision making to select further "promising" structures (non-observed or virtual) to be calculated, aimed at finding the optimum structure – the GM: Fig. 1A – with as few DFT or DFTB calculations as possible.

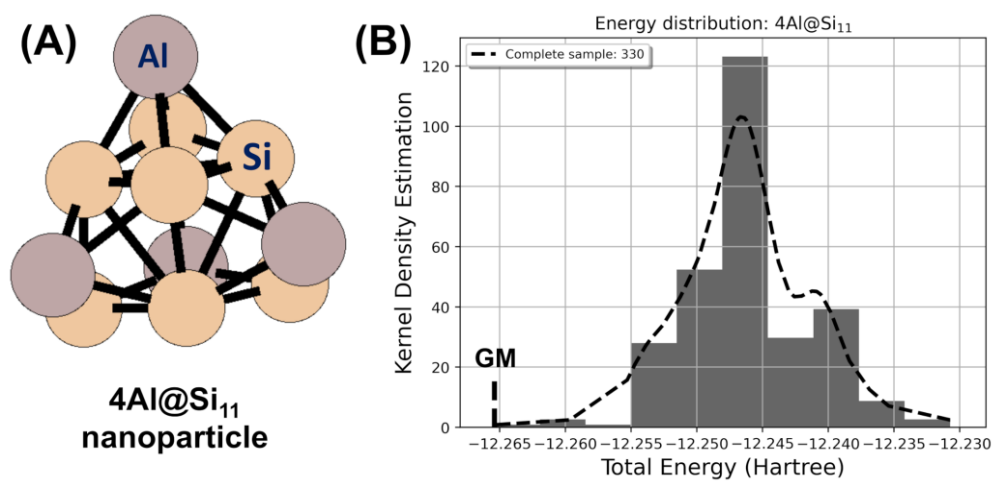

**Figure 1.** (A) Putative global minimum structure (GM) of $4Al@Si_{11}$ found by DFTB and active learning. (B) DFTB energy distribution (in Hartree) of the 330 $4Al@Si_{11}$ homotops or isomers.

This paper is organized as follows: the next section introduces the methods employed in this work, providing a brief overview of active learning, the concepts of classical and quantum kernels as well as classical and quantum ML set up. Section 3 discusses the results of our QAL experiments and compares them with those obtained from classical approaches. Finally, Section 4 summarizes and concludes the work.

## 2. Methods

### 2.1. Quantum active learning

QAL uses supervised QML algorithms to make decisions for the next structures of the doped material (e.g.: nanoparticle) to be calculated[20]. As more data is obtained in each QAL cycle, the QML predictions are improved, increasing the chance of finding the global minimum with as few data or new calculations as possible. Fig. 2 presents the QAL algorithm developed in the present work and

implemented in the QMLMaterial[26] code. The proposed QAL includes following the main componenents:

(A) *Structures of doped materials*: The dataset is initialized using a database containing a small number of doped configurations (denoted as N), which are calculated using methods such as DFT or DFTB[25]. This dataset is then updated during the learning process. At this moment, the structural descriptor such as MBTR[29] (many body tensor representatio) or SOAP[28, 30] (smooth overlap of atomic positions) for the isomers of the nanoparticle is provided (i) for the observed structures (N) – $\boldsymbol{x}^{(i)} = (x_1^{(i)}, \ldots, x_k^{(i)})$, $\forall\ i = 1,\ldots, N$ – and for the unexplored ones or virtual space ($N_{virtual}^k$) – $\boldsymbol{x}^{(j)} = (x_1^{(j)}, \ldots, x_k^{(j)})$, $\forall\ j = 1,\ldots, N_{virtual}^k$. For this work, the MBTR[29] descriptor was used for each $4Al@Si_{11}$ homotop.

(B) Quantum machine learning[23] or *decision-making (the agent)*: In each iteration, a QML regression model is applied to the *N* observed doped structures. Based on either prediction or exploitation, the agent selects the next set of doped nanoparticle structures from the unexplored space $N_{virtual}^k$, denoted as $N_{selected}^{k+1}$ to be calculated in the next step.

(C) *New computations*: The selected set of doped materials, i.e., $N_{selected}^{k+1}$ new structures, are optimized and their total energies are obtained after local optimizations of the structures. The database is then updated ($N = N + N_{selected}^{k+1}$) and the algorithm returns to step (A) and the cycles (*k*) continue until the number of cycles ($N_{cycles}$) set up by the user in QMLMaterial input is satisfied.

Finally, note that by using GPR (or any classical machine learning, ML, model) in step B of Fig. 2 instead of QGPR, classical AL[26] is then defined.

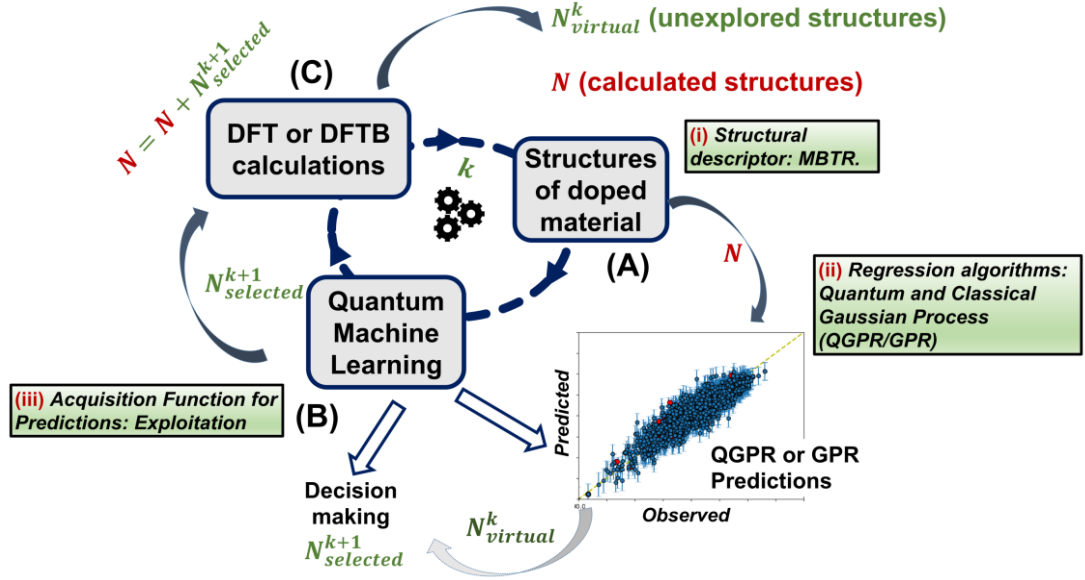

**Figure 2.** Artificial intelligence (AI) workflow based on quantum and classical active learning (QAL and AL) for optimum experimental design. This is implemented in QMLMaterial software using sQUlearn quantum framework.

### 2.2 Classical kernels

Kernel based methods for regression – as in classical GP – are within a model widely employed in supervised learning. It requires a prior's covariance that has to be specified by passing a kernel object. In this work, for classical GP[28], we used two kernels: (1) Dot Product plus White kernel (kernel1 = DotProduct + White) and constant kernel (C) with the radial basis function (RBF) kernel (kernel2 = C*RBF), as implemented in the scikit-learn[31] library.

The DotProduct kernel is given by Eq. 1 where $\sigma_0^2$ is a parameter that controls the inhomogeneity of the kernel.

$$k_{DotProduct}(x_i, x_j) = \sigma_0^2 + \boldsymbol{x_i} \cdot \boldsymbol{x_j} \quad (1)$$

The use of the White kernel is to better estimate the noise level of the data, and is calculated from Eq. 2, where $\nu$ is the noise level parameter (variance).

$$k_{White}(x_i, x_j) = \nu \delta_{ij} \quad (2)$$

The RBF kernel is given by Eq. 1 and is parameterized by the length-scale parameter ($l$).

$$k_{RBF}(x_i, x_j) = exp\left(-\frac{d(x_i,x_j)^2}{2l^2}\right) \quad (3)$$

$x_i$ and $x_j$ are vectors or the descriptors that describe the 4Al@Si$_{11}$ doped distributions (or homotops) configurations $i$ and $j$. $d(x_i, x_j)^2$ is the square of the difference between $x_i$ and $x_j$.

From that, $k_{RBF}$ and $k_{DotProduct}$ kernels are represented by a matrix whose dimension takes into account the number of 4Al@Si$_{11}$ homotops in the training and testing set during the GP fitting and prediction, respectively (step B in Fig. 2).

### 2.3 Quantum kernels

Quantum Kernels (QK) allow QML models to be trained for new inferences in the unexplored space ($N_{virtual}^k$) for new discovery. The idea is to find patterns by transforming data in a vector form (or features) into a high dimensional feature space, $F$. The data mapping from the original input space H to $F$ is done by a feature map $\varphi$: H → $F$. The inner products of the features vectors $\varphi(x_i)$ can access the feature space and a function $k_{FQK}$ of two data points or vectors, $x_i$, $x_j$, is a fidelity quantum kernel[9] (FQK), $k_{FQK}$, Eq. 4.

$$k_{FQK}(x_i, x_j) = \langle \varphi(x_i) | \varphi(x_j) \rangle \quad (4)$$

From Eq. 4 we can observe that $k_{FQK}(x_i, x_j)$ – which depends on the feature map, $F$ – forms an $N \times N$ matrix, in which N is the number of 4Al@Si$_{11}$ doped configurations (e.g. $N=20$) present in the training database.

Another type of quantum kernel is the projected[8] one (PQK), which is based on k-particle reduced density matrices (k-RDMs) measurements (defined in Eq. 5) and was obtained by the sQUlearn quantum computing framework that is interfaced in QMLMaterial. In Eq. 5, $\gamma$ is a real and positive hyperparameter; $\rho_k(x_i) = $ is the 1-RDM for qubit k: the partial trace of the quantum state $\rho_k(x_i)$ over all qubits except for the k-th one. Eq. 5 resembles the RBF kernel (Eq. 1), but in PQK kernel the vectors (or descriptors) $x_i$ and $x_j$ are encoded in the quantum circuit or feature map (e.g.: Fig. 3).

$$k_{PQK}(x_i, x_j) = \exp\left(-\gamma \sum_{k,P}\{tr[P\rho_k(x_i)] - tr[P\rho_k(x_j)]\}^2\right) \quad (5)$$

In Fig. 3 the YZ_CX[18] and HighDim[17] 4 qubits quantum circuits (QC), as implemented in sQUlearn[9], are shown.

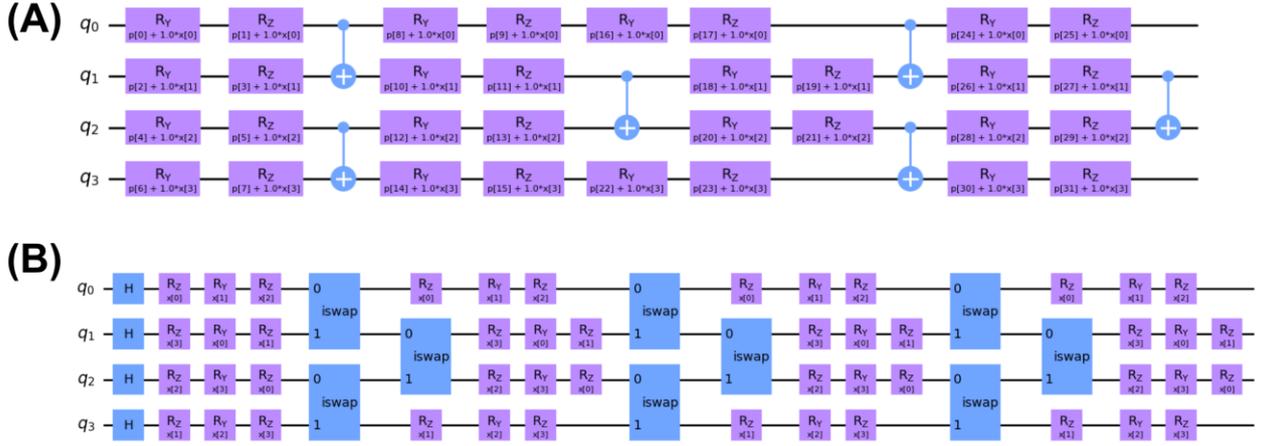

**Figure 3**. (A) YZ_CX and (B) HighDim quantum circuits (QC) as implemented in sQUlearn library. These QCs were built with 4 mapped features or qubits (q0, q1, …, q3). The number of repeated circuits (reps) is equal to 4.

### 2.5 Quantum and classical machine learning set up

The regressions used in QAL are obtained from QGPR[9] and GPR[28] models analytically. The QGPR supervised learning regression algorithm (using different QC encoding to obtain the fidelity quantum kernel, FQK, and projected quantum kernel, PQK) is available in the current version of QMLMaterial, which was developed in Python3.x[32] and uses the scikit-learn[31] library for classical GPR and sQUlearn for quantum computing and QGPR (quantum). In the design loop, the QGPR and GPR regression fit is made on 95 % of the observed data (the training set) and tested on 5 % of it (the test set). This training and testing configuration has been done in other AL studies Ref[4].

Tab. 1 presents the mean absolute error (MAE) for the training and testing set of 4Al@$Si_{11}$ obtained by different QAL and AL iterations. Also, the set up used to obtain the optimum hyperparameters is presented. They were obtained by grid search for a fixed 4Al@$Si_{11}$ energy database.

The QGPR and GPR hyperparameters that resulted in the smallest mean absolute error (MAE) for the training and the testing set are shown in Tab. 1. The optimum hyperparameters found were used subsequently in the AL (classical) study, as shown in Tab. 1. Further information can be found in the documentation of the scikit-learn library.

**Table 1.** Hyperparameters used in this work as defined in the scikit-learn and sQUlearn libraries for the 4Al@Si$_{11}$ nanoparticle. 95 % of the data were randomly chosen for the training set and 5 % for testing for 4Al@Si$_{11}$, as defined during the QAL cycles using exploitation for decision making. The cost function used to evaluate the quality of the regression was the mean absolute error (MAE) and it is presented for different data sizes obtained by QAL. "Hyper." means: hyperparameters used in the QGPR (using projected quantum kernel, PQK, or fidelity quantum kernel, FQK) or classical GPR (radial basis function, RBF, kernel) regression. The MAE is in Hatree $\times$ 10$^{-2}$. The MBTR descriptor was used and PCA was applied, allowing a dimensionality reduction of 4.

| Hyper. | MAE train | MAE test | MAE train | MAE test | MAE train | MAE test |
|---|---|---|---|---|---|---|
| | 20 data | | ~100 data | | ~200 data | |
| featureMap = HighDim; qkernel1 = PQK; σ=0.0001 | 0.0067 | 0.0197 | 0.0031 | 0.0629 | 0.0028 | 0.0611 |
| featureMap = YZ_CX; qkernel2 = PQK; σ=0.0001 | 0.0071 | 0.0156 | 0.0033 | 0.0645 | 0.0031 | 0.0897 |
| kernel1 = DotProduct + White, σ_ bounds[1.0, (10$^{-3}$-10$^{3}$)], noise_level_bounds[10.0, (10$^{-3}$-10$^{3}$)]; α = 1.0 | 0.1374 | 0.1407 | 0.3642 | 0.5137 | 0.3335 | 0.3502 |
| Kernel2 = Constant*RBF, constant_value_bounds[1.0, ($10^{-3}$- $10^3$)], length_scale_bounds [10.0, ($10^{-2}$- $10^2$)]; α = 1.0 | 0.0308 | 67.9609 | 0.0788 | 0.1815 | 0.0528 | 0.0819 |

The GPR and QGPR hyperparameters used during the QAL – as defined in scikit-learn and sQUlearn) that resulted in the smallest mean absolute error (MAE) for the training and the testing set are shown in Tab. 2.

Tab.2 shows the MAE results (for different data size during QAL) of the Projected Quantum kernel[8] (PQK) hyperparameters, from sQULearn library, using two different feature maps: HighDim[17] (Fig. X) and YZ_CX[18] (Fig. Y). For both feature maps and the PQK kernel the optimum σ hyperparameter found was $10^{-4}$ (Tab. 2). σ is the hyperparameter for the regularization strength. During the QAL cycles the σ value was fixed. More details about the HighDim and YZ_CX[18] feature map and σ can be found in the sQUlearn library[9].

The σ parameter in the Dot Product (DotProduct) kernel controls its inhomogeneity and the σ_bounds are from $10^{-3}$ to $10^3$ (*σ_bounds[1.0, ($10^{-3}$- $10^3$)]*), Tab. 2. The noise level parameter in the White kernel (White) controls its variance and its range is: $10^{-3}$ to $10^3$ (*noise_level_bounds[10.0, ($10^{-3}$– $10^3$)]*). In this way the *σ_bound* parameter from DotProduct and the *noise_level* one from White kernel are automatically optimized during the GPR fitting. The combination of two kernels is represented as "DotProduct + White". The Constant kernel (Constant) is used in the classical GPR model with the constant value bounded in the range: $10^{-3}$ and $10^3$ (*constant_value_bounds[1.0, ($10^{-3}$– $10^3$)]*), Tab. 2. This multiplies the Radial Basis Function (RBF) kernel whose length scale bounds are: $10^{-2}$ and $10^2$ (*length_scale_bounds[10.0, ($10^{-2}$- $10^2$)]*). This means that *constant_value* parameter from Constant kernel and the *length_scale* from RBF are automatically optimized during the GPR fitting. The combination of two kernels is represented as "ConstantKernel*RBF". Finally, the α parameter in the GPR models are added to the diagonal of both classical kernel matrices ("DotProduct + White" and "ConstantKernel*RBF") during the fitting. Further information about the kernels can be found in Eq. 1, 2 and 3 and in the documentation of the scikit-learn library[31]. The GP kernel hyperparameters are automatically optimized in the scikit-learn library when the AL evolves and more data is obtained. The α parameter is kept fixed.

### 3. Results and Discussion

The QAL results were obtained for an $Si_{11}$ nanoparticle doped with Al ($4Al@Si_{11}$). The data base was obtained from Ref[2] (the proper Al-Al[33] and Al-Si[2] DFTB parameters was developed for this system using FASP[34] code) and has 330 $4Al@Si_{11}$ structures or homotops calculated by DFTB using the DFTB+ program[35]. Ten independent runs were taken into account and for each one a different distribution of the initial $4Al@Si_{11}$ structure was obtained randomly to define the initial data base with 20 homotops with energies equal or smaller than -12.2400 Hartree, making sure the lower energy structures (local minima) are far away from the global minimum, Fig. 1. More details about the $4Al@Si_{11}$ system and the classical AL method for structural determination using classical AL can be found in Refs[2].

### 3.1. Application to $4Al@Si_{11}$ with 4 qubits in the quantum circuits

In this study the MBTR[36] descriptor was used to describe the $4Al@Si_{11}$ homotops. From that, dimensionality reduction by principal component analysis[7] (PCA) was used. The 4 most representative PCA components (PCA=4) was used and, from that, 4 qubits were considered to build the quantum circuits YC_ZX[18] and HighDim[17]. Also, for classical AL with GPR, the same PCA configuration used in QAL was used, for comparison purposes. The QAL and AL started with 20 random data where all $4Al@Si_{11}$ homotops had energies equal to or higher than -12.2400 Hartree, allowing a scenario far away from the putative global minimum (Fig. 1).

Fig. 4 shows the results of the average total DFTB energy of 4AL@Si$_{11}$ as a function of the number of new DFTB local optimization experiments indicated by the QAL agent for 10 independent runs using 4 qubits to build the quantum kernels and the QGPR models. The QAL with QGPR-YZ_CX-PQK presented the best performance when compared to all other QAL methods. The QAL with QGPR-YZ_CX-PQK performed better in minimizing the 4Al@Si$_{11}$ structures around 140 new local optimization. The classical AL with GPR-kernel2 presented the best overall performance (black circle curves), as shown in Fig. 4. Quite surprisingly, the AL with GPR-kernel1 presented the worst performance (even when compared to QAL) in minimizing the energy of the 4Al@Si11 doped configurations.

The QAL with quantum kernel PQK, when compared to QAL with FQK quantum kernel, presented better performance in minimizing energies of the 4Al@Si$_{11}$ doped configurations – blue and pink curves in Fig. 4. As indicated by the yellow and green curves in Fig. 4, both QAL with FQK kernel – but different feature maps (YZ_CX and HighDim) – presented similar performance. The QAL with QGPR-YC_CX-FQK (yellow curve) performed slightly better after 260 new calculations (i.e.: local optimizations) than QAL with QGPR-HighDim (yellow curve).

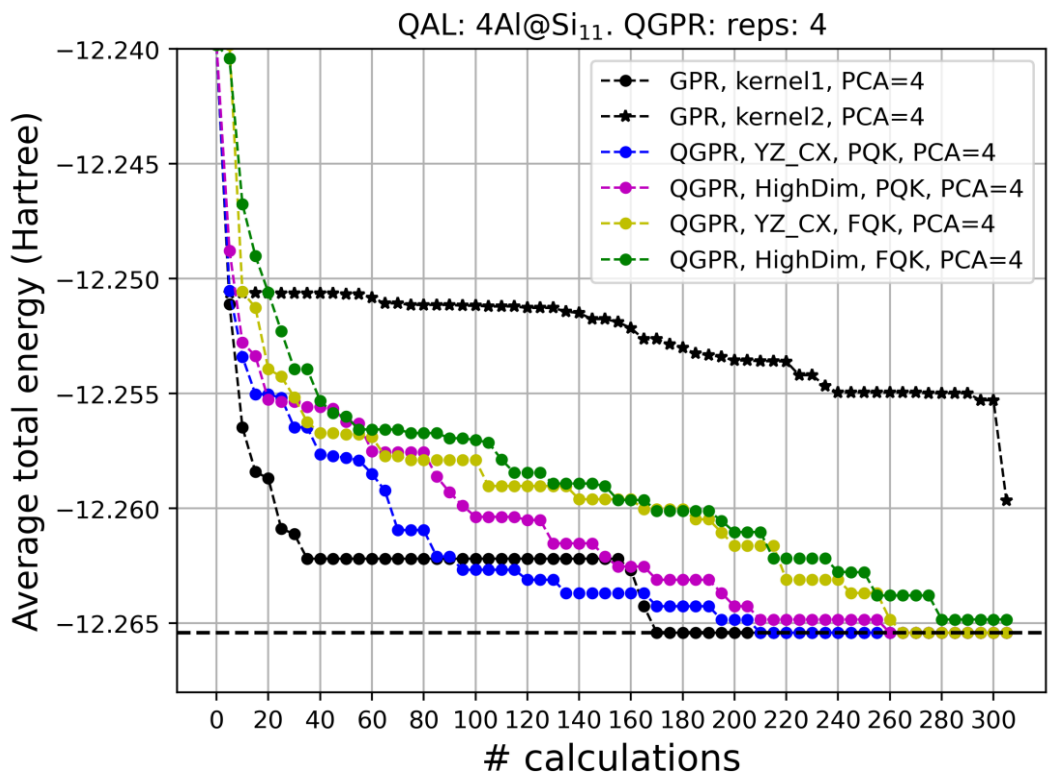

**Figure 4.** Average total energy (in Hartree) obtained by DFTB (in Hartree) for $4Al@Si_{11}$ obtained by 10 QAL and AL independent runs as a function of new calculations (with $N_{cycles} = 60$ and $N_{selected} = 5$). GPR: Classical Gaussian process (GPR) regression with DotProduct and White kernel combination – kernel1 – and radial basis function (RBF) – kernel2. QGPR: Quantum Gaussian process regression with XFeatureMap (X=HighDim, YZ_CX) and Projected Quantum Kernel (PQK) or Fidelity Quantum Kernel (FQK). PCA=X: principal component analysis with X main components chosen from MBTR descriptor. "reps": number of times the quantum circuit is decomposed.

The QC parameters in the QK used to build the QGPR models during the QAL steps (as more data is obtained) are kept constant while the classical kernels used to create GP models have the hyperparameters – as shown in Tab. 1 – automatically optimized. This automatic hyperparameterization of the classical kernels might be one of the reasons the classical AL outperforms the QAL. Of course, further investigations are needed. Even so, the QAL methods, mainly those using GPR-PQK, were able to find the GM.

### 3.2. Application to $4Al@Si_{11}$ with 8 qubits in the quantum circuits

Now we investigate the effect of qubits dimension in the QAL performance for automatic structural determination of $4Al@Si_{11}$ (Fig. 1). The MBTR[36] descriptor was used to represent each $4Al@Si_{11}$ homotops. PCA[7] was taken into account where the 8 most representative PCA components

(PC=8) were used to create the QGPR and GPR models for QAl and AL, respectively. From that, 8 qubits were considered to build the quantum circuits (YZ_CX[18] and HighDim[17]) to build the quantum kernels for QGPR: FQK and PQK. In order to have a challenging search, the QAL and AL methods started with a database with 20 random $4Al@S_{11}$ homotops (or doped Al distributions) with energies equal or higher than -12.2400 Hartree – as was done in the QAL study using 4 qubits.

The results of QAL and AL can be found in Fig. 5. The average total energy (in Hartree) was obtained for 10 independent runs and is expressed as a function of the number of new calculations (i.e.: local optimizations by DFTB). The QAL results for QGPR-YZ_CX-PQK (blue curve in Fig. 5) presented better results with 8 qubits (PCA=8) than with 4 qubits (PCA=4), blue curve in Fig. 4. Also, using PCA=8 for the AL with GPR-kernel1 improved considerably the global search when compared to GPR-kernel1 with PCA=4 (Fig. 4). The putative global minimum (Fig. 1) for the AL with GPR-kernel1 was found (for all independent runs) in just 40 new calculations, as shown in Fig. 5. After 100 new calculations the QAL curve for QGPR-YZ_CX-PQK (blue curve in Fig. 5) stays closer to the AL curve with GPR-kernel1 (black circles curve, Fig. 5). This indicates an improvement of QAL with QGPR-YZ_CX-PQK due to the larger number of qubits: 8.

The AL with GPR-kernel2 and PCA=8 (black star curve in Fig. 5) underperformed the search for the putative global minimum of $4Al@Si11$ as the one with PCA=4 (black star curve in Fig. 4). In all cases the AL with GPR-kernel2 underperformed all AL and QAL results independent of the PCA dimension used as descriptor for the classical GPR.

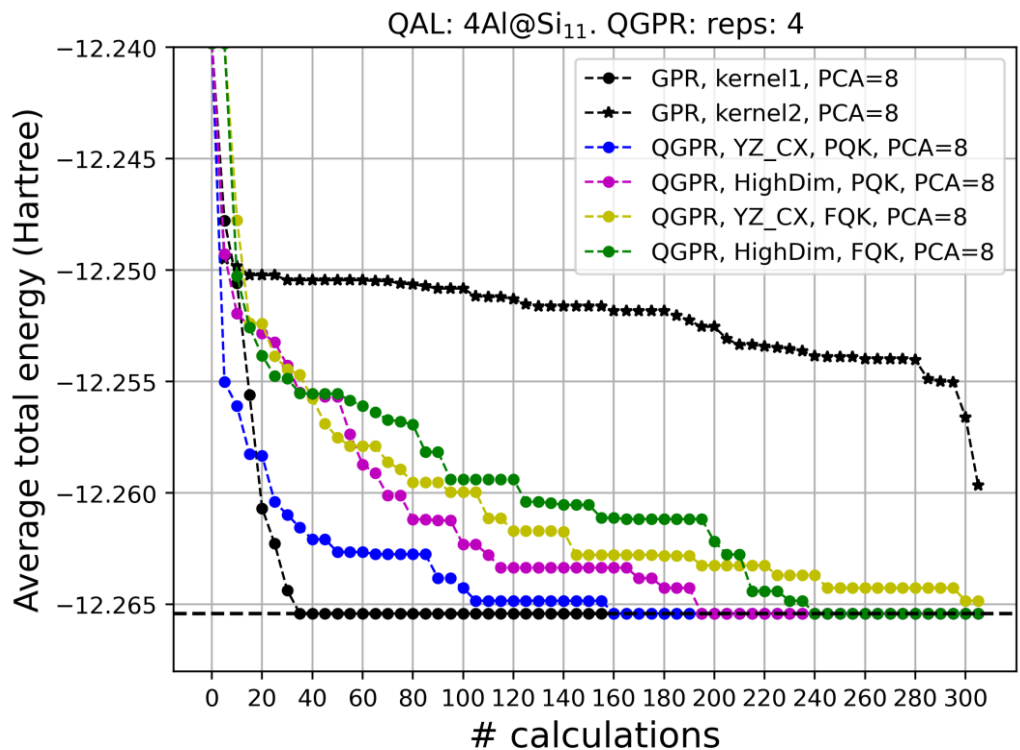

**Figure 5.** Average total energy (in Hartree) obtained by DFTB (in Hartree) for $4Al@Si_{11}$ obtained by 10 QAL and AL independent runs as a function of new calculations (with $N_{cycles} = 60$ and $N_{selected} = 5$). GPR: Classical Gaussian process regression with DotProduct and WhiteKernel combination – kernel1 – and radial basis function (RBF), – kernel2. QGPR: Quantum Gaussian process regression with XFeatureMap (X=HighDim, YZ_CX) and Projected Quantum Kernel (PQK) or Fidelity Quantum Kernel (FQK). PCA=X: principal component analysis with X main components chosen from MBTR descriptor. "reps": number of times the quantum circuit is decomposed.

During the QAL cycles the QC parameters in the QK used to build the QGPR models are constant. On the contrary, the classical kernels used to create GP models have the hyperparameters automatically optimized in the scikit-learn library, as mentioned in Tab. 1. This automatic hyperparameterization of the classical kernels might be one of the reasons for the classical AL performance compared with QAL. Even so, the QAL methods, mainly those with GPR-PQK, were able to find the GM.

4. **Conclusions**

Artificial intelligence (AI) methods based on active learning (AL) have been shown to be efficient for decision-making in small datasets, as in the case of automatic structural determination of point defects[2,3]. In this work we explored the novel meeting of AI and quantum computing (quantum artificial intelligence) by developing a quantum active learning (QAL) method for automatic structural determination of doped nanoparticles using quantum machine learning regression (QML) algorithms. The method was implemented in QMLMaterial software[26] using the sQUlearn quantum computing framework[9] and applied to the global optimization of defects in $Si_{11}$ nanoparticle due to 4 Al doping the Si sites ($4Al@Si_{11}$). The QAL uses the QPGR[23] algorithm created by two kernels: fidelity quantum kernel[9] (FQK) and projected quantum kernel[8] (PQK). The quantum kernels were obtained by using the YZ_CX[18] and HighDim[17] feature maps.

Quantum kernels (QK) and quantum circuits (QC) for data encoding were explored in this QAL study for finding the global minimum (GM) of $4Al@Si_{11}$. Also, different QC dimensions (with 4 and 8 qubits) were explored. Our results showed that QAL with GPR-YZ_CX-PQK presented the best performance in finding the GM in 10 independent runs of QMLMaterial. When 4 qubits are used in the QC to build the QGPR model, the classical AL and QAL are more competitive in finding the 4Al@Si11 GM. On the contrary, for 8 qubits the AL was much better. In all cases, the QAL found the putative GM of $4Al@Si_{11}$, showing that the proposed QAL is feasible for automatic structural determination of point-defect materials. Further QAL development, applications and investigations are necessary to understand the important factors for quantum improvement, such as data roughness[20] and QC design[22].

Finally, the QMLMaterial program (written in Python3.x) now supports QML algorithms that can be explored with different quantum circuits for data encoding for automatic structural determination. This AI software is currently under development with the aim of providing an efficient automatic strutural determination of nanoparticles and solids, important to provide insights into atomic level of materials with defects with as few DFT or DFTB calculations as possible. This quantum artificial intelligence (AI) technology has the potential of finding optimum solutions, within the uncharted chemical space – i.e.: structural or isomers space – in materials science and in other branches of chemistry. So, as QMLMaterial uses quantum information for building QML models and quantum chemistry methods (e.g. DFT or DFTB) to compute homotops energies, a "QQ" methodology is defined: quantum information (Q) for inferences and quantum calculations (Q) of molecules.

**Acknowledgments**

The support of the Brazilian agencies: Fundação de Amparo à Pesquisa do Espírito Santo (FAPES), Conselho Nacional para o Desenvolvimento Científico e Tecnológico (CNPq) and Coordenação de Aperfeiçoamento de Pessoal de Ensino Superior (CAPES) are gratefully acknowledged. Work supported by the National Research Council of Canada, Applications of Quantum Computing program and by the Natural Sciences and Engineering Research Council of Canada, Discovery Grant (RGPIN-2019-03976).

**Author Contributions**

Conceptualization: MPL, MN, LBH, HZH, DG, CS, DRS. Funding acquisition: MPL, DS, CS. Methodology: MPL, MS, HZH, DS. Project administration: MPL, DRS. Resources: MPL, DRS. Software: MPL. Supervision: MPL, MN, DRS, DG, CS. Validation: MPL, MN, LBH, HZH, DG. Writing original draft: MPL. Writing – review & editing: MPL, MN, LBH, HZH, DG, CS, DRS.